\documentclass{article}

\usepackage{amsmath,amssymb}
\usepackage{graphicx}
\usepackage{arxiv}

\DeclareMathOperator{\Var}{Var}

\DeclareMathOperator{\Bin}{Bin}

\title{Adaptive stopping and forecasting of screening, rescreening and randomisation in multicentre clinical trials}

\author{
Volodymyr Anisimov \quad Thomas Henderson\\[0.5em]
{\small Data Science, Center for Design \& Analysis, Amgen, London, UK}\\
{\small E-mail: \texttt{vanisimo@amgen.com},
\texttt{2tomhenderson@gmail.com}}
}
\date{}

\renewcommand{\headeright}{}
\renewcommand{\undertitle}{}
\renewcommand{\shorttitle}{Adaptive stopping and forecasting of screening and randomisation in clinical trials}

\begin{document}
\maketitle

\begin{abstract}

In randomised clinical trials, patients typically undergo one or more screening stages before randomisation. These stages
complicate recruitment planning for trials as patients may be randomised after admissions to screening have been halted.
Stopping the admission of new patients to screening too late may lead to over-enrolment, whereby more patients
than planned are randomised, increasing costs, unnecessary patient exposure and the time required to complete recruitment.
Conversely, stopping too early may result in failure to achieve the planned sample size.
Predictions of the stopping time should therefore account for patients currently undergoing screening and the number of these patients who are ultimately expected to be randomised.

We present a novel statistical framework for jointly modelling screening and randomisation and for determining an adaptive stopping time
to stop admitting new patients to screening before the planned randomisation target is reached. The aim is to reduce over-enrolment
while maintaining a high probability of achieving the target sample size. We represent the screening and randomisation pathway as a multistage process
with stage-specific durations and transition probabilities estimated from accumulated trial data.
The model yields the distribution of the number of patients who will be randomised from the current screening pipeline and,
consequently, the probability of reaching the planned randomisation target. If this probability exceeds a specified threshold,
such as 95\% or 99\%, admission of new patients to screening should be stopped. Otherwise, the model is updated as additional data accrue
and is used to forecast patient arrivals, screening outcomes, randomisations and the future stopping time.

The framework can incorporate scheduled site initiations and future closures through evolving stochastic processes,
thereby supporting forecasts at global, regional and country levels.
Illustrative simulations show that the approach can reduce over-enrolment substantially while preserving a high probability
of reaching the target sample size.


\end{abstract}

\section{Introduction}

Forecasting patient arrivals, the screening pipeline and the times at which operational targets will be reached is an important challenge in large multicentre clinical trials. Such trials often aim to enrol substantial numbers of participants in order to achieve the sample size required for testing the prespecified scientific hypotheses.

An early and commonly used approach modelled patient arrivals as Poisson processes with fixed site-specific rates; see, for example, \cite{senn97}. In practice, however, enrolment rates vary substantially between sites. To represent this heterogeneity, Anisimov and Fedorov \cite{anisimov-fedorov-2007} (see also \cite{an11a}) introduced and popularised the Poisson--gamma (PG) model. In this model, patient arrivals at each site follow a Poisson process, while the site-specific rates are drawn from a gamma distribution.

The PG model \cite{anisimov-fedorov-2007,an11a} has subsequently become one of the most widely used and robust frameworks for forecasting enrolment and has been described in the literature as both ``one of the most popular techniques'' and an ``industry standard'' model. It has also been extended in several directions; see, for example, \cite{an20}. Reviews of alternative approaches are provided in \cite{an16b,barnard10,heitjan-ge-ying-2015}.

Real-world trials, however, typically include one or more screening stages that patients must complete before randomisation, and the structure of the screening pathway is trial-specific. In this paper, we consider trials with screening and rescreening before randomisation. Our aim is to provide practical guidance on stopping the admission of new patients to screening before the randomisation target is reached. This can reduce over-enrolment caused by patients who remain in the screening pipeline at the stopping time and are subsequently randomised, while maintaining a high probability of achieving the target.

It is also often necessary to consider randomisation targets for particular countries or regions. Country-level forecasts can be used to estimate the time required to meet minimum recruitment requirements and to coordinate enrolment schedules so that these requirements are satisfied by the time the overall randomisation target is reached.

An approach to the joint modelling and forecasting of arrival and screening processes, including screening and pre-screening stages and an early stopping rule, was proposed by the first author at the Conference on Advances in Statistics \cite{an25_ICAS}. Extended results for that screening pathway were presented at the 21st ASMDA Conference, held in Piraeus, Greece, on 10--13 June 2025, and are reported in \cite{anisimov-gortzak-2025-asmda}.

In this paper, we consider the more widely used pathway comprising screening followed, where applicable, by rescreening.
The principal contribution is a framework that introduces explicit screening stages between patient arrival and randomisation
and can be adapted to different screening pathways given suitable probability functions. The distribution of the number of patients randomised
after completing these stages is used to derive an operational stopping rule. This rule is combined with forecasts of future arrivals
to predict the earliest monitoring time at which the probability threshold to stop admitting new patients to screening is met.

We also extend the future-arrival model to account for finite site recruitment windows, which can affect the timing and volume of patient arrivals and
subsequent randomisations. We treat these recruitment windows as predetermined, and derive the corresponding site contributions
to model future randomisations under a proposed schedule, supporting forecasts at global, country and regional levels.

The remainder of the paper is organised as follows. We first describe the screening and rescreening pathway and formulate the corresponding mathematical model. We then derive the distribution of the number of randomisations among patients currently in screening or rescreening and extend the model to patients who will arrive in the future. Finally, we present the forecasting and stopping procedure and quantify the resulting risk and expected magnitude of over-enrolment.

\section{Structure of the screening pathway}

\subsection{Screening, rescreening and randomisation}

\begin{figure}[htbp]
    \centering
    \includegraphics[width=0.75\textwidth]{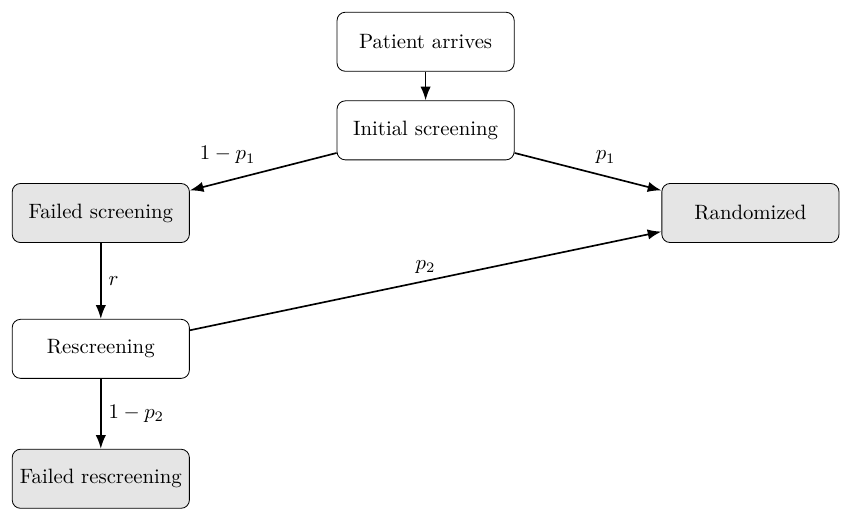}
    \caption{Patient flow through the screening and rescreening pathway, with the associated transition probabilities}\label{fig:screening-logistics}
\end{figure}

Consider a multicentre clinical trial in which patients complete a screening process, with an associated delay, before being randomised into the trial. We focus on a screening/rescreening pathway in which patients may be randomised through either of two routes. They may be randomised directly after an initial screening period of length \( \Delta_1 \), or, following an initial screening failure, they may enter a rescreening period of length \( \Delta_2 \) after a delay of length \( D \). Figure~\ref{fig:screening-logistics} illustrates this pathway.

\subsection{Forecasting at an interim time}

At an interim time \( t_1 \), the numbers of patients currently in screening, \( n_1 \), and rescreening, \( n_2 \), and the additional randomisations required to reach the target, \( L \), are known. The forecast must therefore account for two groups: patients currently in screening or rescreening and patients arriving after \( t_1 \).

\section{Randomisation from the screening pool}\label{Sec3}

To model screening and randomisation at interim time \( t_1 \),
we use information on the patients' progression through the screening stages. Specifically, we consider the number of patients already randomised, the numbers currently in initial screening and rescreening, and the transition probabilities at each stage. These parameters are treated as fixed in the derivations, but in practice they are estimated from interim data once sufficient information has accumulated, as described below.

Table~\ref{tab:screening-pool-notation} defines the observed counts, stage-specific probabilities and process durations. The probabilities and durations are assumed known for the following derivations. The number of additional randomisations required to reach the target is \( L = R - R(t_1) \).

\medskip
\noindent\begin{minipage}{\textwidth}
    \centering
    \refstepcounter{table}
    \label{tab:screening-pool-notation}
    Table~\thetable: Notation for forecasting randomisation from the current screening pool.\par\medskip
    \renewcommand{\arraystretch}{1.2}
    \begin{tabular}{p{0.16\textwidth}p{0.77\textwidth}}
        \hline
        \textbf{Symbol} & \textbf{Definition} \\
        \hline
        \( t_1 \) & Interim forecasting time \\
        \( R \) & Planned randomisation target \\
        \( R(t_1) \) & Number of patients already randomised by \( t_1 \) \\
        \( L \) & Additional randomisations required, \( R - R(t_1) \) \\
        \( n_1 \) & Number of patients currently in initial screening \\
        \( n_2 \) & Number of patients currently in rescreening \\
        \( p_1 \) & Probability of randomisation directly after initial screening \\
        \( r \) & Probability of entering rescreening following initial screening failure \\
        \( p_2 \) & Probability of randomisation for a patient in rescreening \\
        \( \Delta_1 \) & Duration of initial screening \\
        \( D \) & Delay between initial screening and rescreening \\
        \( \Delta_2 \) & Duration of rescreening \\
        \hline
    \end{tabular}
\end{minipage}
\medskip

Using this information, we aim to predict how many patients already in screening or rescreening will eventually be randomised and when these randomisations are expected to occur.

For \( n_1 \) patients currently in initial screening, let \( a_j \) denote the completion times in the interval \( [t_1, t_1 + \Delta_1] \). Similarly, for \( n_2 \) patients currently in rescreening, let \( v_j \) denote the completion times in \( [t_1, t_1 + \Delta_2] \).
Under homogeneous Poisson arrivals, conditional on the number of patients in that stage their entry times are uniformly distributed over the corresponding preceding stage-duration. With fixed stage durations, their completion times are thus uniform between the interim time and the interim time plus the stage duration.
For forecasts based on stage counts alone, uniform remaining completion times provide an approximation when entries to the relevant stages are approximately stable. If individual entry dates are available, the corresponding completion times can be used directly under the fixed-duration model.

\section{Model parameters and estimation from interim data}

The screening and rescreening parameters should be estimated from accumulated data. There are two main groups of parameters. The first describes the structure of the screening pathway, and the second describes the rate at which patients arrive at each site. These components can be estimated separately: the pathway parameters are required to analyse patients already observed in screening or rescreening, whereas the arrival-rate parameters are required to forecast future patients. Both groups are needed to characterise future randomisations.

\subsection{Estimation of screening and rescreening parameters}

For the initial screening stage, let \( m_1 \) be the number of patients who have completed initial screening by \( t_1 \), and let \( s_1 \) be the number of patients who were randomised out of these \( m_1 \) patients directly after initial screening. Let \( m_r \) be the number of patients who entered rescreening after failing initial screening. For the rescreening stage, let \( m_2 \) be the number of patients who have completed rescreening by \( t_1 \), and let \( s_2 \) be the number who were successfully randomised after rescreening. The transition probabilities are then estimated by
\[
    \hat{p}_1 = \frac{s_1}{m_1}, \quad \hat{r} = \frac{m_r}{m_1 - s_1}, \quad \hat{p}_2 = \frac{s_2}{m_2}.
\]

If the relevant sample sizes are small, or if limited trial data have accumulated for a particular stage, information can be pooled across suitable sites within the same country or region to estimate the model parameters. Alternatively, estimates may be borrowed from comparable studies or obtained using Bayesian methods.

The stage durations can also be estimated from patient-level data. Denote the corresponding empirical estimates by
\begin{align*}
    \hat{\Delta}_1 & = \text{mean initial screening duration}, \\
    \hat{D} & = \text{mean delay between screen failure and rescreening entry}, \\
    \hat{\Delta}_2 & = \text{mean rescreening duration}. \\
\end{align*}

In many studies, however, the screening and rescreening windows are specified in the protocol. In such cases, the duration parameters can be treated as fixed constants rather than estimated directly from the data.

\subsection{Site-level arrival parameters}
\label{sec:site-arrival-parameters}

To forecast randomisations among patients who have not yet arrived, we must model the patient-arrival process. For each site \( j \),
suppose that, at interim time \( t_1 \), the site has been active for \( w_j \) days and that \( k_j \) patients have entered initial screening during this period.

The site-level arrival process is modelled as a Poisson--gamma process with a site-specific rate \( \lambda_j \). This rate has a gamma prior with shape parameter \( \alpha_j\) and rate parameter \( \beta_j \), with
mean \( \alpha_j/\beta_j \) and variance \( \alpha_j/\beta_j^2 \).

For a site \( j \) that is already active at time \( t_1 \), Bayesian updating gives a gamma posterior distribution:
\( \lambda_j \sim \mathrm{Ga}(\alpha_j + k_j, \beta_j + w_j) \).
This posterior better reflects the site's observed performance. Estimates for sites with substantial accrued data are driven mainly by their own observations, whereas sites with limited history borrow more information from the prior distribution.

For sites that have not yet opened at \( t_1 \), the arrival rate can be represented by the prior or predictive distribution
\( \lambda_j \sim \mathrm{Ga}(\alpha_j, \beta_j) \),
where \( (\alpha_j, \beta_j) \) may be estimated from comparable sites.

\section{Eventual randomisations from the current screening pool}

To determine an early stopping time that reduces over-enrolment, we first quantify the probability of reaching the target using only patients already observed in the trial. Specifically, we derive the distribution of the number of randomisations among the \( n_1 \) patients in initial screening and the \( n_2 \) patients in rescreening. This distribution yields a stopping rule: if the probability that these patients will provide enough randomisations to reach the target exceeds a chosen threshold \( q \), admission of new patients to screening should stop; otherwise, it should continue.

Let \( X_\infty \) denote the final number of randomisations among patients currently in screening or rescreening.

Using the structure of the screening pathway, \( X_\infty \) can be represented as the sum of two random variables where we account for two cases: a patient currently in initial screening can be randomised either directly after initial screening or after rescreening. Thus, the following equality holds in distribution:
$$ 
    X_\infty \overset{d}{=} \mathrm{Bin}(n_1, p_1 + (1 - p_1) r p_2) + \mathrm{Bin}(n_2, p_2)
$$ 
where the two binomial random variables on the right-hand side are independent.

For a suitably high threshold \( q \), the decision rule is
$$ 
    \mathbb{P}(X_\infty \geq L) \geq q
$$ 
where \( L \) is the number of additional randomisations required to reach the target. If this condition holds, admission of new patients to screening should be stopped.

Here, \( q \) is the required probability of reaching the target using the observed screening pool. To limit the risk of under-enrolment, \( q \) should be high; for example, \( q = 0.99 \). A lower value reduces over-enrolment but increases the risk of failing to reach the target. A higher value reduces the risk of under-enrolment, although it may prolong screening and modestly increase over-enrolment. Because under-enrolment should be minimised, a high threshold is generally appropriate.
The calculations below indicate that increasing \( q \) produces only a relatively small increase in over-enrolment.

The distribution of \( X_\infty \) can also be used to estimate the probability of exceeding specified levels of over-enrolment by varying the value of \( L \).

\section{Randomisations from the current screening pool by time \( t \)}

At interim time \( t_1 \), suppose that \( n_1 \) patients are in initial screening and \( n_2 \) are in rescreening. We use the uniform completion time assumption as described in Section \ref{Sec3}.
Thus, for a patient in initial screening, the probability of completing initial screening by time \( t \) is

\[
    \alpha_1(t) = \frac{t - t_1}{\Delta_1} \wedge 1 \vee 0.
\]

Similarly, the probability that a patient completes the full screening and rescreening pathway by time \( t \), including the delay \( D \) and rescreening duration \( \Delta_2 \), is
\[
    \alpha_2(t) = \frac{t - t_1 - D - \Delta_2}{\Delta_1} \wedge 1 \vee 0.
\]

Finally, for a patient in rescreening at time \( t_1 \), the probability of completing rescreening by time \( t \) is
\[
    \gamma(t) = \frac{t - t_1}{\Delta_2} \wedge 1 \vee 0.
\]

We now incorporate the success probabilities for each stage. A patient in initial screening can be randomised by time \( t \) either directly after initial screening or after rescreening. Direct randomisation occurs with probability \( \alpha_1(t) p_1 \), because the patient must both complete initial screening by time \( t \) and pass that stage.

For randomisation after rescreening, the patient must have entered early enough to complete the entire screening/rescreening pathway by time \( t \), failed initial screening, entered rescreening and passed rescreening. This occurs with probability \( \alpha_2(t) (1 - p_1) r p_2 \). Therefore, the overall probability that a patient currently in initial screening is randomised by time \( t \) is
\[
    A(t) = \alpha_1(t) p_1 + \alpha_2(t) (1 - p_1) r p_2.
\]

For a patient currently in rescreening, the corresponding probability of randomisation by time \( t \) is \( \gamma(t) p_2 \). Let \( X(t) \) denote the number of patients randomised by time \( t \) among those currently in initial screening or rescreening. Then

\[
    X(t) \overset{d}{=} \mathrm{Bin}\left( n_1, A(t) \right) + \mathrm{Bin}\left( n_2, \gamma(t) p_2 \right),
\]
where the two terms account for patients in initial screening and rescreening, respectively.

The corresponding mean and variance are
\begin{align*}
    \mathbb{E}[X(t)] & = n_1 A(t) + n_2 \gamma(t) p_2, \\
    \Var(X(t)) & = n_1 A(t) (1 - A(t)) + n_2 \gamma(t) p_2 (1 - \gamma(t) p_2).
\end{align*}

The full distribution of \( X(t) \) can be obtained by convolving the two binomial distributions.

As \( t \rightarrow \infty \),
$    \alpha_1(t) \rightarrow 1, ~~ \alpha_2(t) \rightarrow 1, ~~ \gamma(t) \rightarrow 1.$

Hence, \( X(t) \) converges in distribution to \( X_\infty \).

\section{Stopping rule and over-enrolment risks}
\subsection{Probability threshold for stopping screening admissions}

Because the distribution of \( X(t) \) is available, we can formulate strategies for deciding when to stop admitting new patients to screening.

Consider a future time \( s \) at which the randomisation target has not yet been reached and a decision must be made about whether to stop admitting new patients.
Let \( R \) be the planned randomisation target, and let \( \Omega_s \) denote the interim information available at time \( s \). This information includes the number of patients already randomised, \( R(s) \), and the numbers currently in screening and rescreening, \( n_1(s) \) and \( n_2(s) \), respectively. Define the number of additional randomisations required to reach the target by \( L(s) = R - R(s) \). If admission of new patients is stopped at time \( s \), let \( W_\infty(s) \mid \Omega_s \) denote the eventual number of randomisations among patients already in the screening/rescreening pipeline. Thus, \( W_\infty(s) \) is the time-\( s \) update of \( X_\infty \), based on \( n_1(s) \) and \( n_2(s) \). Conditional on \( \Omega_s \),

\[
    W_\infty(s) \mid \Omega_s \overset{d}{=} \mathrm{Bin}\left( n_1(s), p_1 + (1 - p_1) r p_2 \right)
    + \mathrm{Bin}\left(n_2(s), p_2 \right).
\]
Admission of new patients to screening should therefore stop at time \( s \) if
$$ 
   \mathbb{P}\bigl( W_\infty(s) \geq L(s) \mid \Omega_s \bigr) \geq q
$$ 
for a chosen probability threshold \( q \), such as 95\% or 99\%. Otherwise, admission of new patients to screening should continue.

\subsection{Over-enrolment risk}

If admission of new patients is stopped on day \( s \), define the resulting over-enrolment, conditional on the observed data, by
\[
    Q_s = \max(W_\infty(s) - L(s), 0).
\]

Conditional on \( \Omega_s \), the distribution of \( Q_s \) is determined by the distribution
of \( W_\infty(s) \):
\begin{align*}
    \mathbb{P}(Q_s = k \mid \Omega_s) & = \mathbb{P}(W_\infty(s) = L(s) + k \mid \Omega_s), \qquad k \geq 1, \\
    \mathbb{P}(Q_s = 0 \mid \Omega_s) & = \mathbb{P}(W_\infty(s) \leq L(s) \mid \Omega_s).
\end{align*}

The expected over-enrolment can be calculated as

\[
    \mathbb{E}\left[ Q_s \mid \Omega_s \right] = \sum_{k = L(s) + 1}^{n_1(s) + n_2(s)} (k - L(s)) \mathbb{P}(W_\infty(s) = k \mid \Omega_s),
\]
or
\[
    \mathbb{E}\left[ Q_s \mid \Omega_s \right] = \sum_{k = L(s) + 1}^{n_1(s) + n_2(s)} \mathbb{P}(W_\infty(s) \geq k \mid \Omega_s).
\]

\section{Randomisation process for patients arriving after the interim time}
\subsection{Arrival times and randomisation probabilities}

If, at interim time \( t_1 \), the probability of reaching the target using only patients already in screening or rescreening is not sufficiently high, admission of new patients should continue. We must then account for patients who enter screening after \( t_1 \) and determine how many of them may be randomised by a future time \( t \).

Let \( Y(t) \) denote the number of patients randomised by time \( t > t_1 \) among those who enter screening after \( t_1 \). To derive the randomisation probability over time for an arriving patient, consider Poisson arrivals with a rate that, conditional on the site rate, is constant. This applies for example to an individual site over an active recruitment time. For \( u < t \), let \( U(u,t) \) denote the number of patients who arrive during the interval \( [u,t] \). Conditional on \( U(u,t)=n \), the arrival times \( y_1,\ldots,y_n \) are independent and uniformly distributed over \( [u,t] \). For patient \( i \), let
\[
    Z_i \sim \mathrm{Bernoulli}(p_1), ~~ R_i \sim \mathrm{Bernoulli}(r), ~~ G_i \sim \mathrm{Bernoulli}(p_2),
\]
where \( Z_i \) indicates successful initial screening, \( R_i \) indicates entry into rescreening after an initial-screening failure, and \( G_i \) indicates successful rescreening. These variables are assumed to be mutually independent and independent across patients. The indicator that patient \( i \) has been randomised by time \( t \) is therefore
\[
    I_i(u, t) = \mathbf{1}\{ y_i + \Delta_1 \leq t \} Z_i + \mathbf{1}\{ y_i + \Delta_1 + D + \Delta_2 \leq t\} (1 - Z_i) R_i G_i.
\]

Because \( y_i \mid \{ U(u, t) = n \} \sim \mathrm{Unif}[u, t] \), define the probabilities

\begin{align*}
    q_1(u, t) &= \left(\frac{t - u - \Delta_1}{t - u}\right) \wedge 1 \vee 0, \\
    q_2(u, t) &= \left(\frac{t - u - \Delta_1 - D - \Delta_2}{t - u}\right) \wedge 1 \vee 0, \\
    h(u, t) &= q_1(u, t)p_1 + q_2(u, t) (1 - p_1) r p_2.
\end{align*}

It follows that \( I_i(u,t) \) is a Bernoulli random variable with success probability \( h(u,t) \). Conditional on \( \{U(u,t)=n\} \), the number of these patients randomised by time \( t \) has a binomial distribution with parameters \( n \) and \( h(u,t) \). This will be applied separately to the aggregation of the active sites and to each site opening in future in the following subsections.

Here, \( q_1(u,t) \) is the probability that a patient whose arrival time is uniformly distributed over \( [u,t] \) completes initial screening by time \( t \). Similarly, \( q_2(u,t) \) is the probability that the patient arrives early enough to complete initial screening, the delay \( D \), and rescreening by time \( t \). Hence, \( h(u,t) \) is the probability that a patient arriving after time \( u \) is randomised by time \( t \).

\subsection{Aggregate contribution from sites active at interim time}

First consider sites that are already active at interim time \( t_1 \), so that their initiation times satisfy \( u_i < t_1 \). For each site \( i \), let \( w_i = t_1 - u_i \) be the length of its active enrolment period and let \( k_i \) be the number of patients who arrived during this period. As described in Section~\ref{sec:site-arrival-parameters}, under the Poisson--gamma model and after Bayesian updating, the posterior rate at site \( i \) is
\[
    \lambda_i \mid \Omega_{t_1} \sim \mathrm{Ga}(\alpha + k_i, \beta + w_i),
\]
where the common prior parameters \( (\alpha,\beta) \), used across all sites, are estimated by maximum likelihood as described in \cite{anisimov-fedorov-2007,an11a}. The aggregate posterior arrival rate across all active sites is
\[
    \hat{\Lambda}_{\mathrm{Act}} = \sum_{u_i < t_1} \lambda_i.
\]

Anisimov and Austin \cite{anisimov-austin-2020} showed that \( \hat{\Lambda}_{\mathrm{Act}} \) can be well approximated by a gamma-distributed random variable \( \Lambda_{\mathrm{Act}} \) with mean and variance
\begin{align*}
    \mathbb{E}[\Lambda_{\mathrm{Act}}] &= \sum_{u_i < t_1} \left( \alpha + k_i \right) / \left( \beta + w_i \right), ~~
    \Var(\Lambda_{\mathrm{Act}}) = \sum_{u_i < t_1} \left( \alpha + k_i \right) / \left( \beta + w_i \right)^2.
\end{align*}

If \( \mathbb{E}[\Lambda_{\mathrm{Act}}] = 0 \), as \( \Lambda_{\mathrm{Act}} \) is non-negative, we have \( \Lambda_{\mathrm{Act}} = 0 \) almost surely. Otherwise, when \( \Var(\Lambda_{\mathrm{Act}}) > 0 \), the distribution of \( \Lambda_{\mathrm{Act}} \) can be approximated by a gamma distribution with parameters \( (A_{\mathrm{Act}},B_{\mathrm{Act}}) \), where

\[
    A_{\mathrm{Act}} = \frac{\left( \mathbb{E}[\Lambda_{\mathrm{Act}}] \right)^2}{\Var(\Lambda_{\mathrm{Act}})}, \quad
     B_{\mathrm{Act}} = \frac{\mathbb{E}[\Lambda_{\mathrm{Act}}]}{\Var(\Lambda_{\mathrm{Act}})}.
\]

For sites active at \( t_1 \), the arrival process over \( [t_1,t] \) is therefore PG with cumulative rate \( (t-t_1) \Lambda_{\mathrm{Act}} \). Each arriving patient is randomised by time \( t \) with aggregate throughput probability \( h(t_1,t) \). By the thinning property of a Poisson process, the cumulative randomisation rate over \( [t_1,t] \) is \( (t-t_1)h(t_1,t)\Lambda_{\mathrm{Act}} \).

\subsection{Randomisations from sites opening in the future}

Consider a new site \( j \) initiated at time \( u_j \geq t_1 \). Let its arrival rate, \( \lambda_j \), be a gamma-distributed random variable with parameters \( (\alpha_j,\beta_j) \). If country- or region-specific information is available, these parameters may be estimated from that information; otherwise, they may be derived from posterior information for active sites or from historical data.

Let \( U_j(u_j,t) \) denote the number of patients arriving at site \( j \) during \( [u_j,t] \), and let \( Y_j(u_j,t) \) be the corresponding number of randomisations at that site during the same interval. By the same argument as above, the conditional distribution of \( Y_j(u_j,t) \), given \( U_j(u_j,t)=n \), is binomial:
\[
  [  Y_j(u_j, t) \mid U_j(u_j, t) = n ] \sim \mathrm{Bin}\left( n, h(u_j, t) \right).
\]

Applying the thinning property of a Poisson process, the unconditional distribution of \( Y_j(u_j,t) \) is Poisson--gamma with cumulative gamma-distributed random rate
\[
    (t - u_j) h(u_j, t) \lambda_j.
\]

\subsection{The global randomisation process}

Combining the contributions from active sites and sites opening in the future, the cumulative random rate for the global randomisation process \( Y(t) \) over \( [t_1,t] \) can be represented as
\[
    \Lambda(t) \overset{d}{=} (t - t_1) h(t_1, t) \Lambda_{\mathrm{Act}} +
    \sum_{t_1 \leq u_j < t} (t - u_j) h(u_j, t) \lambda_j.
\]

The process \( Y(t) \) is therefore represented as an evolving stochastic process; see \cite{an16}. The mean and variance of the global random rate are given by

\begin{align*}
    \mathbb{E}[\Lambda(t)] &= (t - t_1) h(t_1, t) \mathbb{E}[\Lambda_{\mathrm{Act}}] +
    \sum_{t_1 \leq u_j < t} (t - u_j) h(u_j, t) \frac{\alpha_j}{\beta_j}, \\
    \Var(\Lambda(t)) &= (t - t_1)^2 h^2(t_1, t) \Var(\Lambda_{\mathrm{Act}}) +
    \sum_{t_1 \leq u_j < t} (t - u_j)^2 h^2(u_j, t)\frac{\alpha_j}{\beta_j^2}
\end{align*}

\subsection{Special case without rescreening randomisations}

Consider the case where patients can only be randomised through initial screening, so that \( r = 0 \) and
\( p_2 = 0 \). The general framework therefore reduces to a single pathway. In particular, the eventual number of
randomisations from the current screening pool satisfies

\[
X_\infty \overset{d}{=} \Bin(n_1, p_1),
\]

and the number of randomisations by time \( t \) among patients currently in screening satisfies

\[
X(t) \overset{d}{=} \Bin(n_1, \alpha_1(t) p_1).
\]

For patients arriving after the interim time, the probability of randomisation by time \( t \) reduces
to
\[
h(u, t) = q_1(u, t) p_1.
\]

The future arrival distributions thus retain the same structure, but with simplified probabilities.
The stopping rule and over-enrolment calculations apply using the corresponding single-binomial
distribution.

\section{Site closing time and recruitment windows}
\subsection{Site-level recruitment exposure}

Let \( u_j \) denote the opening time of site \( j \) and \( c_j \) the time at which the site stops admitting new patients to initial screening. These are assumed to be predetermined.
Note that all patients admitted before \( c_j \) are allowed to complete the full screening/rescreening pathway, so that the site may continue to contribute randomisations after \( c_j \).
If no recruitment closure is scheduled, \( c_j \) can be set to \( \infty \).

At interim time \( t_1 \) consider for each site the contributions to the forecast. The active exposure at site \( j \) between
\( t_1 \) and a future time \( t \) is

\[
    w_j(t_1, t) = (\min(t, c_j) - \max(t_1, u_j))_+,
\]

where \( (x)_+ = \max(x, 0) \). This represents the number of days during which the site is actively recruiting patients to initial screening between \( t_1 \) and \( t \).
The active recruitment window is zero if the site is not open during the interval \( [t_1, t] \).

Let \( \lambda_j \) be the arrival rate at site \( j \). For a site with observed recruitment data let \( \alpha_j, \beta_j \) denote the posterior parameters, and for sites not yet open let
\( \alpha_j, \beta_j \) denote the prior parameters. The arrival rate is then \( \lambda_j \sim \mathrm{Ga}(\alpha_j, \beta_j) \).
Conditional on this rate, the number of patients entering screening during the active window satisfies
\[
    U_j(t_1,t) \mid \lambda_j
    \sim \mathrm{Pois}\bigl(\lambda_j w_j(t_1,t)\bigr),
\]

where \( U_j(t_1, t) \) denotes the site-level arrival count.

Conditional on \( U_j(t_1, t) = n \), the arrival times are uniformly distributed on
\( [\max(t_1, u_j), \min(t, c_j )] \). For a non-empty active window, the corresponding stage completion and
randomisation probabilities are

\begin{align*}
    q_{1j}(t_1, t) & = \bigl( \frac{t - \max(t_1, u_j) - \Delta_1}{\min(t, c_j) - \max(t_1, u_j)} \bigr) \wedge 1 \vee 0, \\
    q_{2j}(t_1, t) & = \bigl( \frac{t - \max(t_1, u_j) - \Delta_1 - D - \Delta_2}{\min(t, c_j) - \max(t_1, u_j)} \bigr) \wedge 1 \vee 0, \\
    h_j(t_1 ,t) & = q_{1j}(t_1, t)p_1 + q_{2j}(t_1, t) (1 - p_1) r p_2.
\end{align*}

Here \( h_j(t_1, t) \) is the probability that a patient arriving at site \( j \) during the active window is randomised by time \( t \).

If \( w_j(t_1, t) = 0 \), set \( q_{1j}(t_1,t) = q_{2j}(t_1, t) = h_j(t_1, t) = 0 \) to avoid accounting for sites with
zero exposure.

Let \( Y_j(t_1, t) \) denote the number of randomisations by time \( t \) among these new arrivals. Under the existing
independence assumptions and with fixed screening probabilities,
\[
    Y_j(t_1, t) \mid \{ U_j(t_1, t) = n \}
    \sim \mathrm{Bin}\bigl(n, h_j(t_1, t)\bigr).
\]
Equivalently, by Poisson thinning,
\[
    Y_j(t_1, t) \mid \lambda_j
    \sim \mathrm{Pois}\bigl( \lambda_j w_j(t_1, t) h_j(t_1, t) \bigr).
\]

Note that if \( c_j = \infty \), the recruitment window becomes \( [\max(t_1, u_j), t] \), and thus the functions
\( q_{1j}, q_{2j}, h_j \) reduce to the previous definitions.

\subsection{Site-level rates and aggregations}

We now apply the same gamma moment-matching technique as in the open-site model, but first calculate the contribution of each site
using its own opening and closing times. This accounts for sites already active at \( t_1 \), sites opening subsequently, and sites
that stop admitting patients during the forecasting period. A site that has closed screening may still contribute randomisations from patients admitted after \( t_1 \) but before its \( c_j \). Patients already in the
screening pool at \( t_1 \) remain part of \( X(t) \) and are not considered by the future-arrival component.

Define the site-level cumulative rate by
\[
    \Lambda_j(t_1,t) = w_j(t_1, t) h_j(t_1, t) \lambda_j.
\]
Thus, the site-level predictive count is Poisson--gamma, with rate moments
\begin{align*}
    \mathbb{E}[\Lambda_j(t_1, t)] & = w_j(t_1, t) h_j(t_1, t) \frac{\alpha_j}{\beta_j}, \\
    \Var(\Lambda_j(t_1,t)) & = w_j^2 (t_1, t) h_j^2(t_1, t) \frac{\alpha_j}{\beta_j^2}.
\end{align*}

The site opening and closing times and screening probabilities are treated as fixed, and the site-specific arrival rates and
patient processes are assumed independent. Summing over all sites gives
\[
    Y(t) = \sum_j Y_j(t_1, t), \qquad \Lambda(t)=\sum_j \Lambda_j(t_1, t).
\]
Sites with zero future recruitment exposure contribute zero to
these sums. Consequently,
\begin{align*}
    \mathbb{E}[\Lambda(t)] & = \sum_j w_j(t_1,t ) h_j(t_1, t) \frac{\alpha_j}{\beta_j},\\
    \Var(\Lambda(t)) & = \sum_j w_j^2(t_1, t) h_j^2(t_1, t) \frac{\alpha_j}{\beta_j^2}.
\end{align*}

If \( \mathbb{E}[\Lambda(t)] = 0 \), then as \( \Lambda(t) \) is non-negative, \( \Lambda(t) = 0 \) almost surely and consequently  \( Y(t) = 0 \) almost surely. Otherwise, when \( \Var(\Lambda(t)) > 0 \), the distribution of \( \Lambda(t) \) can be approximated by a gamma distribution with parameters \( (A_{\mathrm{future}}(t), B_{\mathrm{future}}(t)) \), where
\[
    A_{\mathrm{future}}(t) = \frac{\bigl( \mathbb{E}[\Lambda(t)] \bigr)^2}{\Var(\Lambda(t))},
    \qquad B_{\mathrm{future}}(t) = \frac{\mathbb{E}[\Lambda(t)]}{\Var(\Lambda(t))}.
\]
This gives a Poisson--gamma approximation for \( Y(t) \). Contributions
from a particular country or region can be obtained by restricting
the sums to its sites. The single-screening special case also applies under site closures,
with \( h_j(t_1, t) = q_{1j}(t_1, t) p_1 \).

\section{Operational forecasting algorithm}

The simulation procedure follows Anisimov and Gortzak \cite{anisimov-gortzak-2025-asmda}. Patient arrivals and subsequent screening outcomes are simulated, the probability of reaching the target using the current pipeline is evaluated, and admission of new patients is stopped once the probability threshold is reached. We adapt this procedure to the screening/rescreening pathway developed in the present paper. The simulation also accommodates site operating periods, including scheduled site openings and closures before the overall stopping time. If the stopping criterion is reached before a scheduled site opening, that site does not begin recruitment.

Specifically, at the end of each day \( s \), the trial state is updated using the number already randomised, \( R(s) \), and the numbers currently in screening and rescreening, \( n_1(s) \) and \( n_2(s) \). The stopping rule is then used to evaluate the conditional probability of reaching the target using only patients already in the pipeline. Admission of new patients to screening stops at time
\[
    s^\star = \min\left\{ s: \mathbb{P}\left( W_\infty(s) \geq R - R(s) \mid \Omega_s \right) \geq q \right\}.
\]

Define the minimum of an empty stopping set as infinity, so that if the stopping criterion is never reached, \( s^\star = \infty \). In forecasts with finite site recruitment windows, simulations where the probability threshold is not reached within the monitoring horizon are recorded separately and their frequency reported.

No new patients are admitted after \( s^\star \), but those already in the screening/rescreening pipeline are allowed to complete the pathway. We obtain the predictive distribution of \( s^\star \) by Monte Carlo simulation conditional on the trial information \( \Omega_{t_1} \). In each replication, we first simulate the remaining outcomes and completion times for the \( n_1(t_1) \) patients in screening and the \( n_2(t_1) \) patients in rescreening at the interim time. Future daily arrivals are generated from a fitted site-level Poisson--gamma model, subject to site initiation and closure times, and each patient follows the screening pathway. At the end of each day \( s \), the trial state is updated and the conditional probability in the stopping rule is evaluated. The first day on which this probability reaches \( q \) is recorded as \( s^\star \). Repeating the procedure over \( M \) independent replications gives the sample \( s^\star_1,\ldots,s^\star_M \), whose empirical distribution approximates the predictive distribution of the screening stopping time.

\subsection{Illustrative simulation}

Following the simulation and presentation approach of Anisimov and Gortzak \cite{anisimov-gortzak-2025-asmda}, we present one realisation of the adaptive stopping procedure for the screening/rescreening pathway. All times and durations are expressed in days. The simulation began with 30 active sites, with 20 additional sites opening one per week starting on day 7. Sites scheduled to open after adaptive stopping did not begin recruitment. In each replication, a single arrival rate for each site was drawn from \( \mathrm{Ga}(1.2,60) \) and retained throughout the replication. Conditional on these rates, daily arrival counts under the site operating schedule are generated as independent Poisson counts, with arrivals after adaptive stopping excluded. The screening parameters were: \
$    p_1 = 0.75, ~~ r = 0.50, ~~ p_2 = 0.70, ~~ \Delta_1 = 14, ~~ D = 7, ~~ \Delta_2 = 14.$

The randomisation target was set to \( R = 100 \), and admission of new patients to screening was stopped on the first day for which the conditional probability of reaching the target was at least \( q = 0.99 \).

\begin{figure}[!htbp]
    \centering
    \includegraphics[width=0.6\textwidth]
        {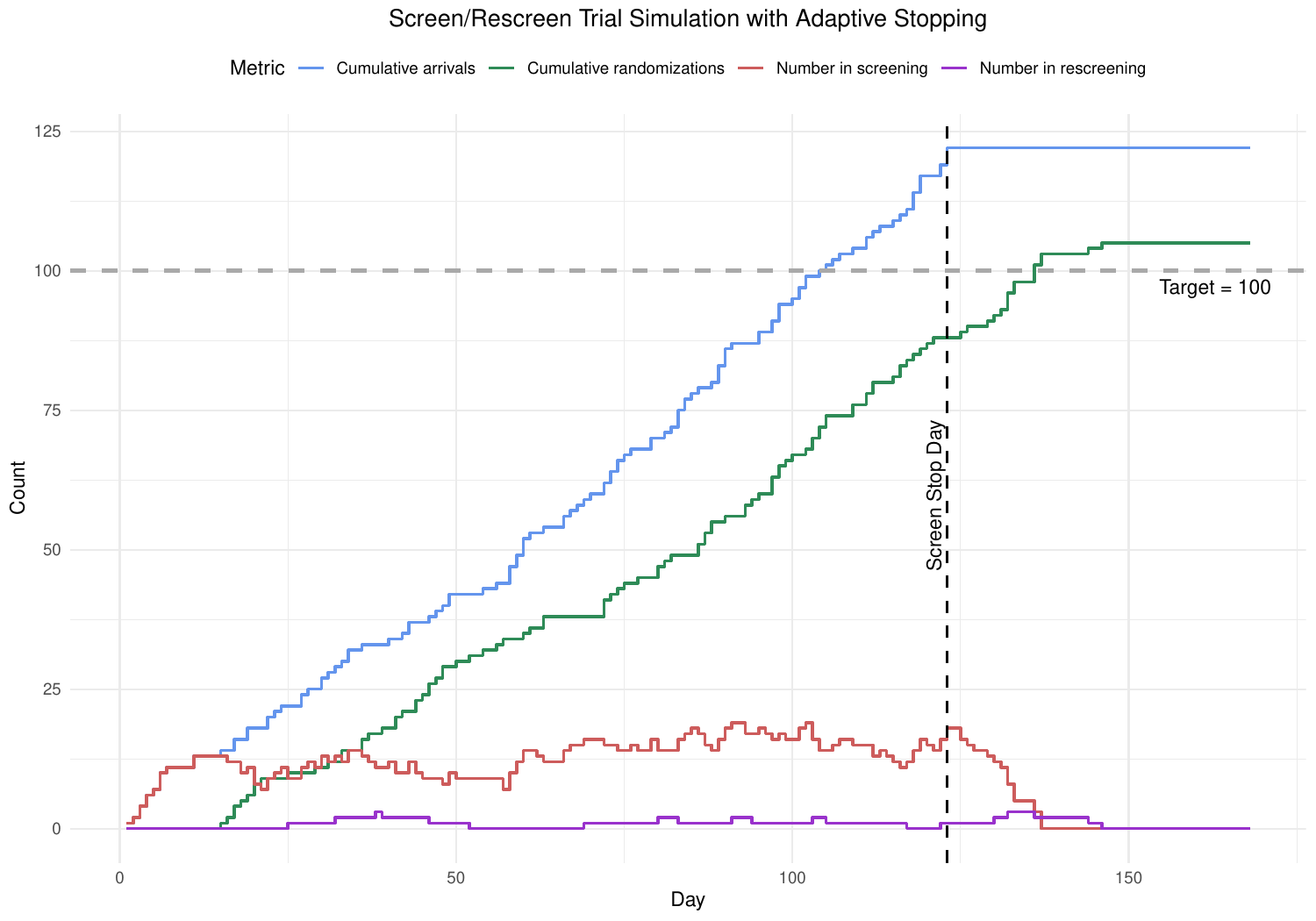}
    \small
    \caption{One simulated realisation under the adaptive stopping rule. The dashed vertical line marks the stopping day, \( s = 123 \). By this time, 88 patients had been randomised, with 18 patients in initial screening and one in rescreening, giving a conditional probability of 0.9914 of reaching the target. The target was reached on day 136, and the final number of randomised patients was 105, corresponding to over-enrolment of five patients.}
    \label{fig:adaptive-screen-rescreen-simulation}
\end{figure}

\begin{figure}[!htbp]
    \centering
    \includegraphics[width=0.6\textwidth]
        {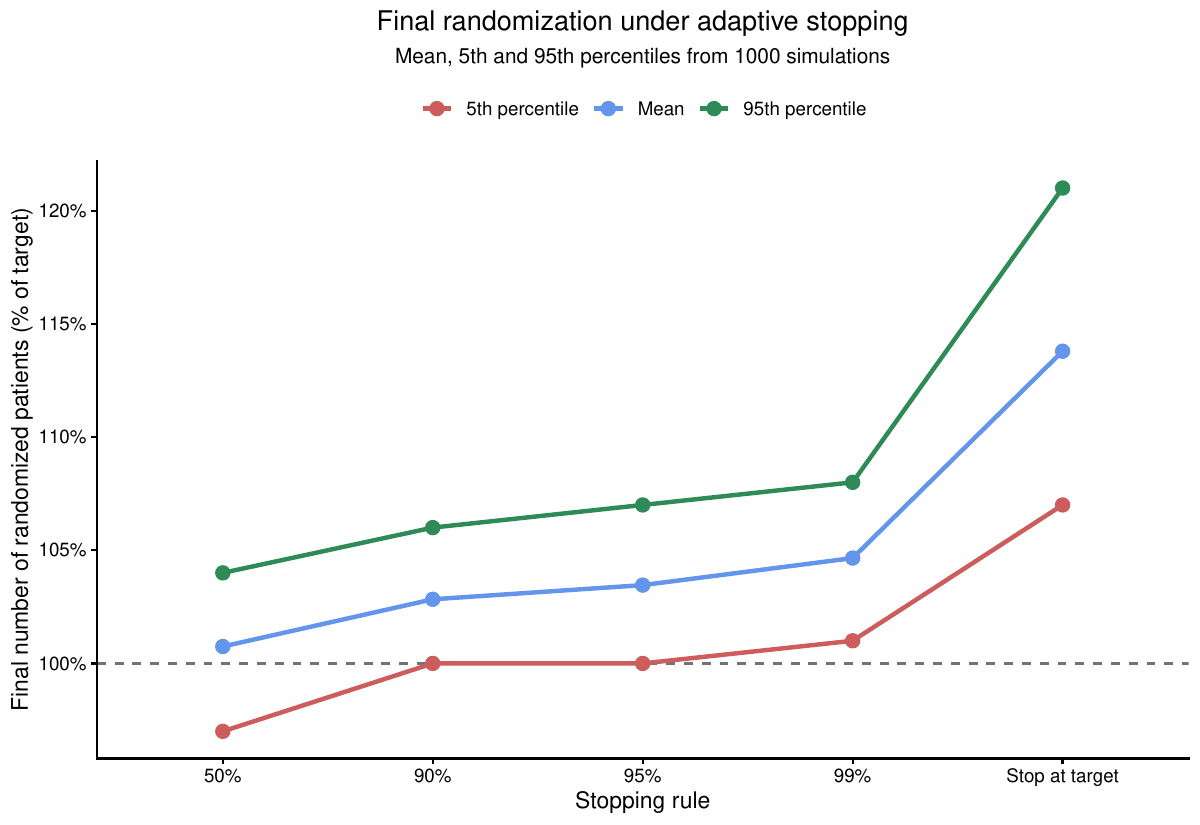}
    \small
    \caption{Final number of randomised patients in simulated trials under stopping thresholds of 50\%, 90\%, 95\% and 99\%, together with the baseline rule that stops admission only when the target is reached. The points show the mean and the 5th and 95th percentiles across 1000 simulations, expressed as percentages of the target \( R = 100 \). The final number includes all patients who completed screening or rescreening after admission of new patients had stopped.}
    \label{fig:adaptive-threshold-enrollment}
\end{figure}

To examine the trade-off between under-enrolment and over-enrolment under different stopping rules, we repeated the simulation 1000 times using the same screening, rescreening and recruitment parameters. We considered probability thresholds of \( q = 0.50,0.90,0.95,0.99 \) and compared the resulting distributions of the final number of randomised patients with a baseline rule under which admission of new patients to screening stops only when the randomisation target \( R \) is reached.
All replications under each policy reached their respective stopping criteria within the 300 day monitoring horizon. The results are shown in Figure~\ref{fig:adaptive-threshold-enrollment}.

Figure~\ref{fig:adaptive-threshold-enrollment} shows that the over-enrolment distributions are relatively similar for the high thresholds of 90\%, 95\% and 99\%, whereas over-enrolment is substantially greater under the baseline rule that waits until the target has been reached.

\subsection{Combining forecasts for current and future patients}

The preceding results concern two disjoint patient groups. The process \( X(t) \) denotes the number of randomisations by time \( t \) among patients already in screening or rescreening at interim time \( t_1 \), whereas \( Y(t) \) denotes the number of randomisations among patients arriving after \( t_1 \). Thus the total number of additional randomisations between \( t_1 \) and \( t \), denoted by \( Z(t) \), is
\[
    Z(t) = X(t) + Y(t).
\]

The cumulative trial total by time \(t\) is
\(R(t_1)+Z(t)\).
Conditional on the interim information and fixed fitted model parameters, \( X(t) \) and \( Y(t) \) are independent under a predetermined admission schedule. Therefore the predictive distribution of \( Z(t) \) can be obtained by convolving the distributions of \( X(t) \) and \( Y(t) \).
Evaluating this distribution over time provides forecasts of the number of randomisations by time \( t \), incorporating both patients observed at the interim analysis
and those expected to arrive afterward.

\section{Conclusion}

This paper presents a novel statistical framework for forecasting randomisation and determining when to stop admitting new patients to screening
in multicentre clinical trials with screening and rescreening, with the aim of reducing over-enrolment. We derived the distribution
of future randomisations from the current screening pipeline and combined it with forecasts of new patient arrivals based on a Poisson--gamma model
that accommodates future site initiations and site closures. The resulting framework enables calculation of the conditional
probability of attaining the randomisation target under different scenarios. The simulation study illustrates how admission of new patients can be
stopped before the target is reached while retaining a high probability of achievement. The present model assumes fixed transition
probabilities and screening durations. Forecasts with closure schedules can be used to inform operational decisions via scenario planning. Future work could involve the extension of this model to include priors for the
screening probabilities and durations.

\section*{Acknowledgements}

The authors are grateful to Matt Austin, the Data Science Team at the Center for Design \& Analysis, and the Clinical Program Operations Team at Amgen for their valuable and continuing discussions on this work.

\end{document}